# 6G Digital Twin Networks: From Theory to Practice

Xingqin Lin[†], Lopamudra Kundu[†], Chris Dick[†], Emeka Obiodu[†], and Todd Mostak[§]

[†]NVIDIA, [§]HEAVY.AI

Emails: {xingqinl, lkundu, cdick, eobiodu}@nvidia.com, todd@heavy.ai

*Abstract—* Digital twin networks (DTNs) are real-time replicas of physical networks. They are emerging as a powerful technology for design, diagnosis, simulation, what-if-analysis, and artificial intelligence (AI)/machine learning (ML) driven real-time optimization and control of the sixth-generation (6G) wireless networks. Despite the great potential of what digital twins can offer for 6G, realizing the desired capabilities of 6G DTNs requires tackling many design aspects including data, models, and interfaces. In this article, we provide an overview of 6G DTNs by presenting prominent use cases and their service requirements, describing a reference architecture, and discussing fundamental design aspects. We also present a real-world example to illustrate how DTNs can be built upon and operated in a real-time reference development platform – Omniverse.

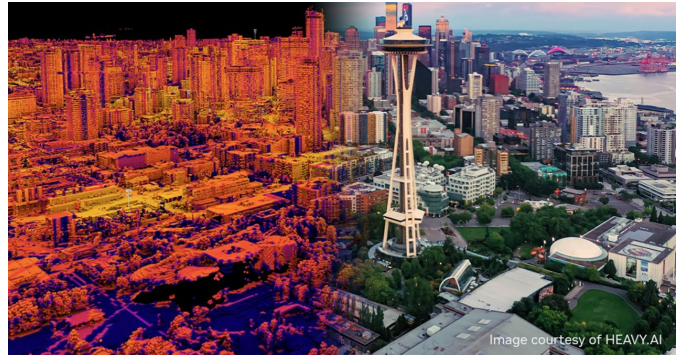

**Figure 1: Digital twins enable virtual replicas of physical reality.**

## I. INTRODUCTION

Digital twins (DTs) are a key part of the industrial metaverse that simulates physical assets such as factories, cities, and transportation systems in the virtual world. With recent advances in software platforms, artificial intelligence (AI)/machine learning (ML), and high-performance computing accelerated by graphics processing units (GPUs), DTs have gained traction in many different areas, such as smart cities, manufacturing, and retail [1]. Meanwhile, wireless networks have become more intricate over time [2]. The complexity of the sixth-generation (6G) wireless networks will grow due to their scale, multi-vendor components, and the need to support diverse use cases [3]. This trend calls for innovative tools and platforms like DTs to facilitate the design, analysis, and operation of wireless networks for 6G and beyond.

A DT network (DTN) is a digital replica of the full life cycle of a physical network. It uses data and models to create a physically accurate network simulation platform, which provides up-to-date network status and predicts future network state [4][5]. It also provides interfaces for interactions with the physical network and network applications/users. Unlike conventional network simulators, the DTN supports two-way communication between the physical network and the virtual twin network to achieve real-time interactive mapping and closed-loop decisions [6].

DTNs will be a foundational 6G technology to realize the vision of a cyber-physical continuum between the physical world and its digital representation. Figure 1 shows an example of creating virtual replicas of physical reality with DT technology. Inspired by the great potential of what DTs can offer, recently there has been much work painting visions of DTNs for 6G [4]-[7]. There are also emerging initiatives in standards bodies which develop initial guidelines for DTNs. The International Telecommunication Union (ITU) radiocommunication sector (ITU-R) is finalizing the report on future technology trends of terrestrial International Mobile Telecommunications (IMT) systems towards 2030 and beyond (aka. 6G), which lists DT as an important use case and discusses technologies to support DTN [8]. The ITU telecommunication standardization sector (ITU-T) released a recommendation that describes requirements and architecture for DTNs [9]. The Institute of Electrical and Electronics Engineers (IEEE) interfacing cyber and physical world working group also proposed a standard series IEEE 2888, e.g., P2888.1 defines sensor interface between cyber (i.e., DT) and physical world [10].

Despite that the existing work presents vivid visions of DTNs for 6G and identifies opportunities, we are at the very beginning of making DTNs a reality for 6G. Realizing the desired capabilities of 6G DTNs requires tackling many design aspects, which have not been adequately addressed in the existing work. This article aims to bridge the gap in the existing literature by not only providing an overview of DTNs for 6G with a focus on key design aspects, but also describing how DTNs can actually be built and operated in Omniverse [11] – a scalable, multi-GPU, real-time reference platform for building and operating metaverse applications.

## II. USE CASES AND REQUIREMENTS

### A. Use Cases

DT applications are destined to become ubiquitous across every industry. Increasingly, before anything is made in the physical world, it will be first simulated in a virtual world and a DT will be created. Furthermore, everything that moves will one day be autonomous, and these autonomous machines will be taught and trained in virtual worlds. For 6G networks and network operations, DTNs can support many use cases including, but not limited to, the examples illustrated in Fig. 2 and described in the sequel.

*Network simulation and planning:* DTNs improve network simulation and planning process using large-scale, physically accurate digital replicas of city blocks. This allows network deployment teams to test and optimize installation and placement of base stations and their configurations in a virtual world first, thereby reducing the expenditure of resources and budget in the physical world. 6G standardization may also exploit DTNs to carry out simulation rather than relying on stochastic channel models and hexagonal deployments. In particular, DTNs with ray tracing are well positioned to simulate the deployment of reconfigurable intelligence surfaces (RISs).

*Network operation and management:* One of the objectives of the softwarization of 6G is the capability to change virtually all aspects of the networks through software, from the physical layer (e.g., site specific optimizations to even bespoke user-specific waveforms) all the way to the core network. Even the topology of the fronthaul network will become software defined. The DTN can be used to create a virtual drive-test to maintain system robustness and resilience, prior to applying the configuration changes in the real network. The DTN can also be used to predict future outages based on network behavior, prompting, if necessary, corrective actions to circumvent the outage. Likewise, the DTN enables the network operators to evaluate multiple scenarios to understand actual customer experience and minimize churn. This can be used for real-time capacity planning for events or to better assess network performance before onboarding new customers.

*Data generation by simulation:* In AI-native 6G networks, access to real-world data is crucial. However, it is not always possible to obtain a suitably rich dataset that reflects diverse set of network operating conditions, e.g., operating environment, weather, and network operational parameters. Synthetic data is annotated information that a simulation running in a DTN generates as an alternative to real-world data. Put another way, synthetic data is created in a cyber sibling of the physical world rather than being measured or collected from the real world. It may be artificial, but synthetic data reflects real-world data in a statistical sense. DTNs capture the precise geometry and material properties of objects in the environment to produce datasets for training AI/ML models used for optimization and network control [12].

*AI training and inference:* AI/ML applied in image and video processing employs models that are trained offline with datasets such as ImageNet, Cityscapes, and the like. Inference engines are deployed with these *a priori* generated model parameters. We anticipate that while this paradigm will be applied in some areas of wireless, it is more likely that online training will play a bigger role in 6G and beyond. One example is deep reinforcement learning (DRL) based closed-loop control. A neural network embedded in the DRL agent will be continually updated in response to the environment. Latency will be important and so training and inference speeds will be critical. Therefore, for DRL based real-time control in 6G, it is essential to utilize a DTN that natively resides on a programmable hardware architecture equipped with hardware accelerators that support training and inference pipelines. In short, faster-than-real-time simulation capabilities of DTNs will be critical for AI-based network optimization and control in real time.

*What-if-analysis:* 6G networks will be dynamic, with network operators being able to modify key network parameters and architecture in the time scale of minutes and hours, rather than months or years. In this case, the DTN is a virtual sandbox that identifies network misconfigurations, link bottlenecks, and security issues to enable prediction of future performance. In many instances, network failure is a rare occurrence, and so it is difficult to produce a large enough dataset to train a neural network. While not feasible in the real network, faults can be deliberately injected into the cyber sibling for the purpose of generating a rich dataset for ML or other data analytics processes. Also, DTNs can enable high-fidelity predictions to reduce randomness in a system, which is a key enabler for ultra-reliable communications.

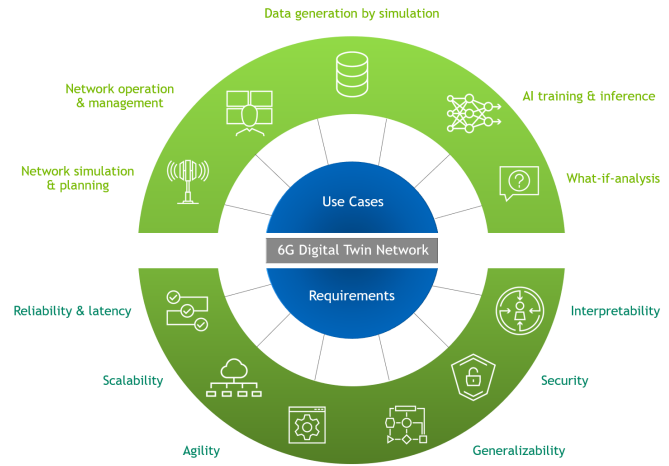

**Figure 2: Use cases and requirements of 6G digital twin networks.**

B. *Requirements*

The foundation of 6G DTNs is built upon a set of design principles shaped by various use case requirements, including reliability & latency, scalability, agility, generalizability, security, and interpretability, as illustrated in Fig. 2.

*Reliability & latency*: Trustworthiness is at the core of the DTN's fundamental characteristics. Rendering high degree of stability throughout the lifecycle of a DTN requires robustness in its operational infrastructure, including proactive handling of latency critical adverse situations involving human error, equipment failure or malicious attacks with real-time response, capability of ensuring robust data collection, storage, modeling, and information exchange between the DTN and other network entities, high degree of availability, and capabilities for disaster recovery (such as backup provision, and ability to restore critical historical states/data points).

*Scalability*: Scale of a DTN can vary over a wide range, depending on the dimension and complexity of its physical counterpart. A physical network may grow or shrink in scale and the DTN should automatically adjust itself. Such flexible scalability is an essential attribute of DTNs across various network domains.

*Agility*: Serving various network applications on-demand requires a DTN to have sufficient flexibility in its functionalities as needed. Ability for cross domain interaction, information exchange and service cooperation between multiple DT entities are some of the important attributes of a DTN towards meeting needs of various operational stages of a physical network.

*Generalizability*: Applicability of DTNs across network equipment of various network applications and topologies requires a certain degree of compatibility. Specifically, data collection, storage, modeling, and interfaces need to be generalizable to support multi-vendor, multi-standard interoperability and ensure backward compatibility between updated and older versions of DTNs.

*Security*: Guarantee of adequate protection from potential attacks as well as defense mechanism for threat prevention should be imprinted in the DNA of a robust DTN. A comprehensive security and privacy consideration for DTN elements including data, models, interactive interfaces, and overall network infrastructure is warranted to ensure integrity of a DTN.

*Interpretability*: Augmenting DTNs with easy-to-use asset management services aids user-friendly management of overall lifecycle of the twin entities like network equipment, links, traffic flows, etc. Ability to observe discernable changes in the functionalities of a DTN through display tools improves interpretability of the DTN's behavior with associated events like instance creation, aggregation, update, and termination.

## III. REFERENCE ARCHITECTURE

In this section, we describe the architectural aspects of 6G DTNs that enable interactive virtual-real mapping and control. Based on the ITU-T recommendation Y.3090 [9], a reference 6G DTN architecture may consist of three layers: 6G physical network layer, 6G twin layer, and 6G network application layer. Figure 3 provides an illustration of the reference 6G DTN architecture.

The 6G physical network layer refers to the target real-world 6G network, consisting of physical network elements and their operating environment. Depending on the need, a 6G DTN may focus on different parts of the target 6G physical network (e.g., a radio cell, a radio access network, a transport network, and a core network) or the entire end-to-end 6G network. The 6G physical network layer exchanges data and control messages with the 6G twin layer.

The 6G twin layer is the core of a 6G DTN. It consists of three domains: data domain, model domain, and management domain. Data domain is a data repository subsystem responsible for collecting data from the 6G physical network in order to build accurate and up-to-date models in the twin. The functions of the data repository subsystem may include collection, storage, service, and management of data. Model domain is a service-mapping subsystem comprising of models to represent the real-world objects in the 6G physical network based on the collected data. There are two types of models: basic and functional. Basic models represent network elements and topology to describe the target physical network. Functional models refer to analytical models for extracting insights from the DTN. Examples include network planning, traffic analysis, fault detection, network emulation, and prediction. Management domain is a DT entity management subsystem responsible for the management function of the 6G twin layer. Examples include model management (e.g., model creation, configuration, update, and monitoring), and security management (e.g., authentication, authorization, encryption, and integrity protection).

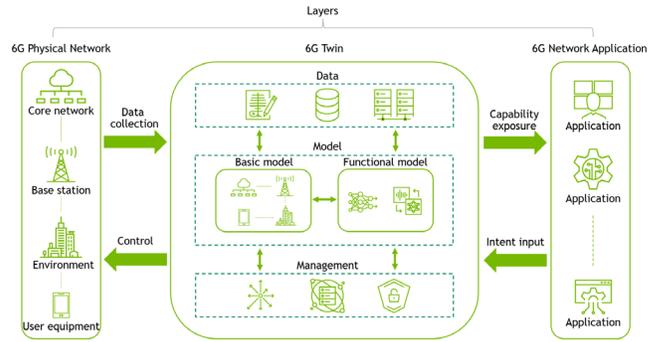

**Figure 3: A reference 6G digital twin network architecture based on ITU-T recommendation Y.3090.**

The 6G network application layer includes various network applications that exploit the capabilities exposed by the 6G twin layer. The network applications provide intent to the 6G twin layer which then emulates the required services and, if needed, sends control messages to the 6G physical network layer. Examples of the network applications include network operations, administration, and maintenance (OAM), network optimization, and network visualization.

There will be diverse 6G DTNs with different capabilities to meet the demands of a wide range of use cases. It is vital to develop common components, generally applicable frameworks, standardized interfaces, and universal platforms and tools to enable interoperability and extensibility of the 6G DTNs. These developments will gradually evolve into unified tools to build the diverse 6G DTNs on universal platforms hosting data, models, and management functions.

## IV. FUNDAMENTAL BUILDING BLOCKS

With a reference 6G DTN architecture in place, the next question is how we can realize it. The inception of a DTN begins with the blending of three fundamental building blocks: data, models, and interfaces. The degree of reliability with which a DTN can mirror its physical counterpart depends on the quality of data upon which its models are built, and the accuracy of the models. Interfaces around a DTN, on the other hand, enable its seamless interaction with the underlying physical network and associated network applications.

### A. Data

Data is the basis for building a 6G DTN. Detailed, up-to-date data enables the construction of high-accuracy models in the DTN. With AI/ML at its core, the efficacy of a 6G DTN relies fundamentally on both the quality as well as the quantity of data collected from different parts of the physical network infrastructure.

*What-to-collect (data type):* In general, the collected dataset needs to be comprehensive enough to constitute a holistic representation of the physical network. At the same time, proper analysis of specific network topology to determine required data type is crucial in order to avoid redundant data collection, which may result in unnecessary cost of compute and storage resources. Some of the data type examples include data representing operational and provisional status of a physical network, network telemetry, performance measurement, data from network services/applications and lifecycle management (LCM) of network entities, and user data. Striking the right balance between data quality and quantity is essential in building a sustainable and trustworthy DTN.

*When-to-collect (time/frequency):* In order to swiftly adjust to the dynamic variation of a live network, the DTN requires constant monitoring, performance reassessment, and timely adjustment. Collection of data to feed a DTN, therefore, cannot be a one-time event. The frequency and pattern of the time series of data collection depend primarily on the data type. For example, data related to faults/events are usually collected in an event-triggered manner, whereas network performance specific data are usually collected at a regular interval, with potential aperiodic trigger enablement depending on pre-defined conditions/intents. Some types of data that statistically vary over shorter periods of time (e.g., port/interface statistics) require higher collection frequency. On the other hand, some other types of data that are critical to network's operational health (e.g., flow status and link failures) need to be gathered in a time sensitive manner.

*How-to-collect (mechanism/tool):* Building a comprehensive DTN requires network data with diverse characteristics, which, in turn, necessitates usage of various data collection mechanisms and tools. The state of physical devices may be collected by the associated sensors that send periodic or event-triggered updates to data collectors. The state of physical sites may be captured by using lasers and drones besides connected sensors. Integrated sensing and communication utilizing terahertz (THz) or optical frequency bands can perform high-resolution sensing, localization and tracking, and 3D imaging to collect data for building and updating the models of wireless propagation environments. Data collection capabilities can be embedded in the 6G physical network to collect status about network elements, topology, traffic, etc. Technologies enabling simultaneous and collaborative data collection from heterogeneous network nodes in a performant and time-synchronized manner are critical in establishing a high-quality data collection process to serve as the solid foundation of an ultra-reliable DTN.

*Where-to-store (repository):* Collection of massive amounts of network data comes with the burden of complex storage requirements. To that end, construction of a unified data repository for myriad network data becomes one of the key prerequisites for creating a successful DTN. The salient features of a hyperconverged data repository include support for heterogeneous database, provision for efficient data services (such as data augmentation and federation, historical data reporting snapshot and rollback, and fast database query), and assurance of high data availability, among others.

*How-to-access (retrieval):* Acceleration is the key for efficient operations (e.g., data ingestion, search, query, and visualization) on the DTN data repository which has large-volume data and fast data streams ingestion from the physical network. To this end, it is crucial to exploit the massive parallel capabilities provided by GPU and the corresponding tools such as Dask to accelerate the three-phase data processing - extract, transform, and load. This is particularly important when the database operations need to be real-time to synchronize the DTN with its physical counterpart, derive lightning-fast data analytics, and achieve immediate closed-loop decisions.

*How-to-maintain (management):* Long term stability, adaptability, and performance of a DTN depend on the efficacy of the underlying data management framework, including the LCM of data collection, storage, maintenance, and retrieval. Data management can either be an integral part of the overall DTN management framework or can be an independent entity in the twin layer. Automation is arguably the most desired feature of data management, along with other capabilities like secure handling, storage, deposition and destruction of sensitive network data (e.g., data related to user privacy), traceability and accurate records of historical data transactions, and guarantee of data integrity (i.e., accuracy, completeness, and consistency).

### B. Models

DTs of a physical network can be modularized at different levels, depending on the use case requirement. For example, DTNs can be of individual physical network domains (like radio access/core/transport network), or of a combination of collocated network entities/functionalities (like DT of cell site/near edge/far edge/cloud), or at the extreme case, a single DTN for the entire end-to-end physical network. The degree of modularity, in turn, determines specific aspects of a DTN. For example, a DTN responsible for a cell site can be operating at a high-level of detail using high-fidelity ray tracing employing multi-bounce specular reflection, 3D diffraction, and diffuse scattering to enable site-specific optimizations in the base station's physical and medium-access control layers. A DTN in the cloud would be responsible for end-to-end service analysis and network operation including resource management and, for example, AI/ML approaches for network slicing.

Arguably, the most challenging design aspect of these multimodal DTNs is their models that represent accurate digital versions of physical entities constituting the real network peers. The generic requirements of a typical DTN model are manifold. The crux of a DTN model meeting the diverse set of requirements is built upon two pillars: physically accurate modeling and its functional manifestation using AI-powered simulation, visualization and control.

*Physically accurate modeling*: High-fidelity models are fundamental building blocks for 6G DTNs. The models need to accurately capture the up-to-date characteristics of their physical counterparts, including geometries, materials, properties, lighting, behaviors, and rules, among others. One key requirement of the models for the 6G DTN is the need of physically accurate simulation of radio wave propagations in true-to-reality environments. This is particularly pertinent considering that 6G will utilize THz technology with ultra-wide signal bandwidth, integrate sensing with radar or lidar technology, use extremely large antennas, and incorporate RIS. Therefore, it is vital to accurately model the effects of radio wave reflection, diffraction, scattering, and multipath propagation in the environment. Traditional software ray tracing techniques may have difficulty in performing efficient, physically accurate simulation of the radio wave propagations in the complex 6G systems. Ray tracing at-scale is extremely computationally intensive. At least two components are required to bring radio frequency (RF) ray tracing at-scale to reality in a DTN: 1) Programmable hardware accelerators supporting the mathematical operations at the heart of ray tracing and 2) optimized libraries for the ray tracing pipeline. An example of a ray tracing hardware acceleration (RTX) are the ray tracing cores found in the many hundreds of streaming multiprocessors that comprise the NVIDIA Ampere class GPU and that are capable of performing many trillions of floating-point math operations per second. There are several ray tracing frameworks available including DX12, Vulkan, and OptiX. Take OptiX for example. Based on compute unified device architecture (CUDA), it is an application framework for achieving optimal ray tracing performance on a GPU that takes advantage of RTX hardware accelerators. OptiX enables the developers to concentrate on wave propagation rather than on low-level ray tracing optimization. Creating the geometry for the DT at interactive rates has become feasible based on capture using instant neural graphics primitives [13].

*AI-powered simulation, visualization, and control*: Interactive virtual-real mapping and control is a key distinguishing characteristic of 6G DTN, compared to the traditional link, system, or network simulators. However, real-time interaction and control is challenging due to the large amount of data, stringent latency requirements, and computation-intensive simulations. AI-powered functions are key to achieve an autonomous feedback loop between the 6G physical network and its DTN. As one example, AI techniques can leverage historical and real-time data to create and evolve models in the 6G twin layer to constantly run simulations for anomaly detection and network optimization, leading to intelligent decision making for the target 6G physical network. AI models can also be used to constantly run predictive "what-if" simulations in the 6G twin layer to determine in advance the consequences of hypothetical actions, which allows for impact analysis of changes, prevention of outage events, and finding optimal settings for future operations. In addition, AI techniques can create photorealistic 3D graphics to visualize the 6G DTN and illustrate findings, facilitating the involvement of humans in the decision-making processes.

## C. Interfaces

Open and standard interfaces between twin layer and physical network layer/network application layer are essential towards proliferating a multi-vendor, interoperable DTN ecosystem. In general, the guiding principles of interface protocols around DTNs include extensibility, backward compatibility, easy usage and accessibility, capability of handling high concurrency of dataflow, and mechanisms to ensure secure and reliable communication pathway between DTNs and other network entities.

*Network-bound interfaces*: As the bridge between a DTN and its underlying physical network, network-bound interfaces provide efficient information exchange between the endpoints, including both control and user plane data. Network-bound interfaces carry various types of network data (e.g., provisional and operational data) collected via different methods such as on-demand, subscription-based, event-triggered, direct measurement, and indirect observation. The interfaces support versatile data requirements from DTN models, resulting in varying data transfer speed requirements ranging from real-time (~ms) to near-real time (~second) and to non-real time (~minute). It is imperative to support management and control signaling through these interfaces for the DTN to configure data collection protocol, time/frequency of collection, and other configurations of network elements in the physical network. Realization of network-bound interfaces using low-latency interconnects is crucial in enabling real-time interactive mapping between the DTN and its physical embodiment. Another important aspect of network-bound interfaces is 'decoupling,' i.e., making the interfaces agnostic of the type of information being transported and specificities of physical entities at its endpoints. Decoupling enables the DTN to use a harmonized interface towards the physical network, without requiring data specific or network entity/topology specific customization.

*Application-bound interfaces*: Delivering requirements (related to network management, optimization, and protocol validation) or intents of network applications to a DTN is fulfilled through application-bound interfaces. Application-bound interfaces support the capability to convey the DTN's features, including common data models and their performance to authenticated, third-party network applications. Speed, latency, and bandwidth requirements for application-bound interfaces are less stringent and less demanding than network-bound interfaces, which makes these interfaces realizable in relatively lightweight ways with reduced complexity. Application-bound interfaces provide a layer of abstraction towards the DTN, implying that the differences between network applications are transparent to the DTN, and the same set of interfaces serves seamless interaction between the endpoints, without requiring application-specific tailoring.

*Intra/inter-DTN interfaces*: Depending on the use case, scalability and flexibility requirements of a DTN drive the degree of modularity, resulting in either a single DTN model or multiple DTN models. For the latter, inter-DTN interfaces enable efficient communication between various DTN models that are either co-located or distributed. For a DTN built upon federated learning-based ML models, both communication between distributed model objects (during training) and aggregation of trained objects can be facilitated through intra-DTN interfaces. While standardization of inter-DTN interfaces between multiple DTN models is crucial for multi-vendor implementation, intra-DTN links between multiple model objects leading to a DTN model ensemble can be proprietary.

## V. OMNIVERSE FOR DIGITAL TWIN NETWORKS

The ambitious goals and complexity of 6G demand DTN implementation frameworks that provide a high-level programming abstraction, large scale modeling, AI/ML training and inference capabilities, tools to interact with the twin through application programming interfaces, and visualization capabilities. Such implementation frameworks enable developers to rapidly implement DTs and let them focus on only the essential complexity of the problem. One such example is the Omniverse platform [11].

Taking advantage of the Omniverse platform, the wireless industry is making progress towards building DTNs. For example, Ericsson has built a city-scale 5G DTN in Omniverse [13]. The 5G DTN helps accurately simulate the interplay between the physical network and the environment for maximum performance and coverage. As another example, HEAVY.AI, a startup that offers an eponymous GPU-accelerated analytics platform that enables users to interactively query, visualize, and power data science workflows over multi-billion record datasets, has recently started building HeavyRF, a network DT application in Omniverse [15].

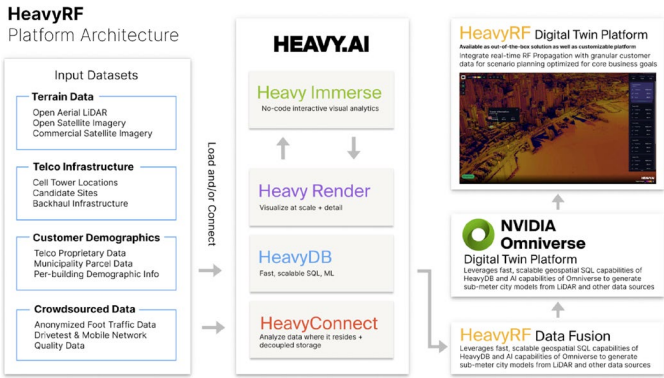

**Figure 4: An example of building digital twins with Omniverse.**

As illustrated in Fig. 4, HeavyRF integrates Omniverse with the HeavyDB structured query language (SQL) backend, leveraging the latter for real-time processing of lidar and other terrain data, as well as executing ray-traced RF propagation simulation itself. Accurate 5G and 6G network planning requires simulation to be performed on unprecedentedly high-resolution terrain data, due to the higher transmitter densities required as well as the high degree of clutter attenuation associated with the use of higher transmission frequencies. HeavyRF can run real-time RF simulations on lidar data at resolutions as granular as 10 cm, which are two orders-of-magnitude higher than the common 10 m plus resolutions used by conventional RF planning solutions for 4G networks, creating true-to-life representations of the physical landscape that a telco operates in. In addition, it can incorporate building-level customer and demographic information to tie the results of network simulation with an operator's key business drivers.

The simulation itself is run directly in the HeavyDB database using the construct of an SQL table function that takes as inputs subqueries representing the terrain, transmitter, and building data, avoiding the overhead that would otherwise be associated with moving these large datasets between servers/software platforms. The use of SQL to orchestrate the simulation also makes it easy to perform inline extract, load, and transform on the input data and various enrichment operations on the simulation output, such as joining the terrain locations and simulated signal strengths to building polygons to measure the minimum, maximum, and average signal power received for each building, as well as capture the principal towers serving each building. HeavyRF optimizes communication between HeavyDB and Omniverse to provide a real-time capability to manipulate network parameters such as tower locations, antenna gain and transmissions frequency and then visualize the results in full 3D. Figure 5 provides a visualization of HeavyRF for network planning and operation. Users can place new towers or change the parameters of existing towers, and see the resultant changes in simulated RF signal across the scene's terrain, as well as rolled up to the building level and tied to key customer metrics.

In summary, Omniverse is a powerful platform for delivering DT solutions. As more and more innovative DT applications are built in Omniverse, we anticipate that it will be a game changer in many facets of wireless research, development, and deployment in the future.

## VI. CONCLUSION

The fast growth of network scale towards 6G and the stringent performance requirements of diverse use cases call for innovative tools and platforms. The unique capabilities of DTNs make them a powerful technology for the design, analysis, diagnosis, simulation, and control of 6G wireless networks. In this article, we have provided an overview of DTNs for 6G and presented a real-world example to illustrate how DTNs can be built and operated in Omniverse. The innovations and applications of DT technology in wireless networks are as vast as the imagination. It will be exciting to see how DT technology will improve lives and transform industries on the path to 6G and beyond.

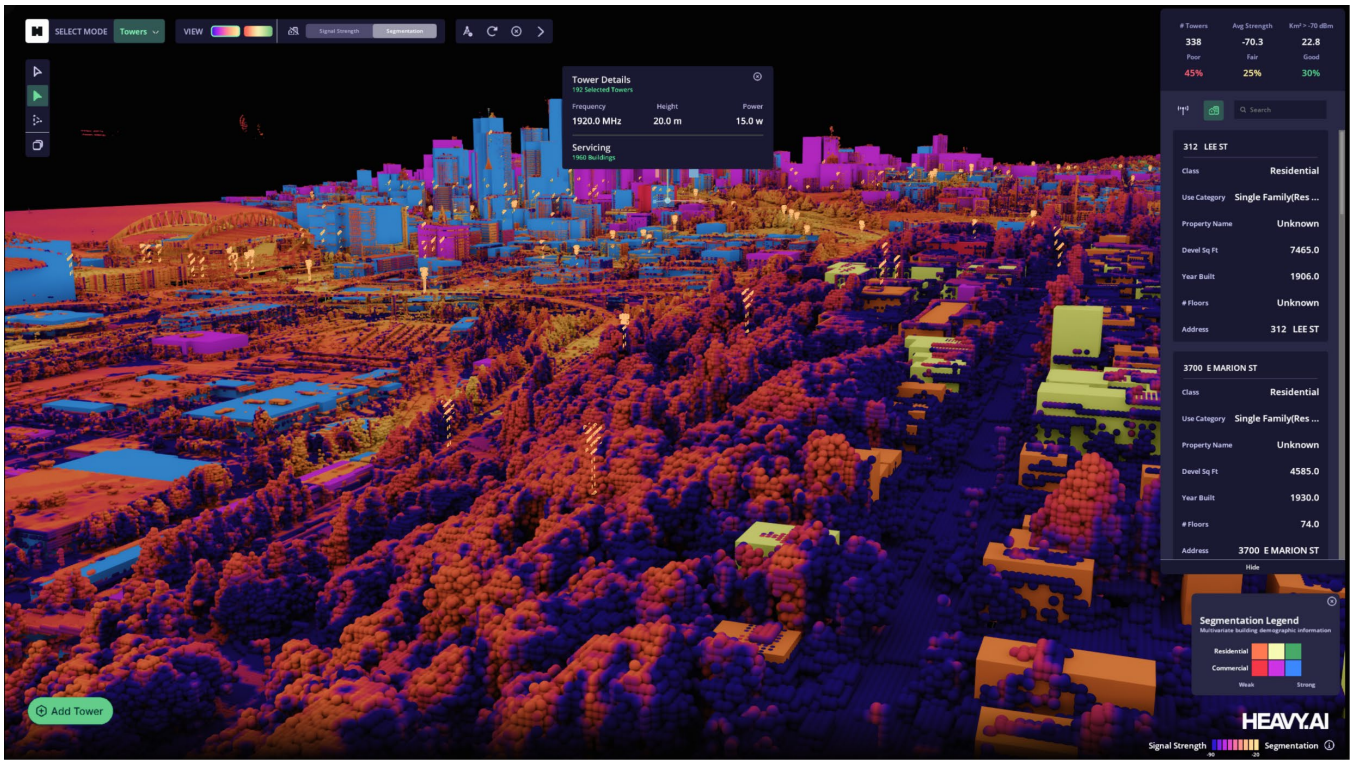

**Figure 5:** A visualization of HeavyRF built with Omniverse for network planning and operation.

BIOGRAPHIES

**Xingqin Lin** is a Senior 3GPP Standards Engineer at NVIDIA.

**Lopamudra Kundu** is a Senior Standards Engineer at NVIDIA.

**Chris Dick** is a Wireless Architect at NVIDIA.

**Emeka Obiodu** is a Global Marketing Manager at NVIDIA.

**Todd Mostak** is the CTO and co-founder of HEAVY.AI.